# Monomer motion in single- and double-stranded DNA coils

J. Tothova, B. Brutovsky, and V. Lisy

*Institute of Physics, P.J. Safarik University,*

*Jesenna 5, 041 54 Kosice, Slovakia*

**Abstract.** The dynamics of flexible polymers in dilute solution is usually described in terms of the pure Rouse or Zimm bead-spring models assuming continuous distribution of the internal relaxation modes. We show that this approach may lead to misleading interpretation of experimental data. The more correct description should come from the joint Rouse-Zimm (RZ) theory that contains the Rouse and Zimm models as limiting cases. The internal modes are discrete with respect to the mode number, and the type of the bead motion changes in the time from the Rouse to Zimm behavior. We demonstrate this interpreting the recent first observation of the kinetics of individual polymer monomers using the fluorescence correlation technique [R. Shusterman *et al.*, Phys. Rev. Lett. **92**, 048303 (2004)]. Optimizing the RZ theory to the data on double- and single-stranded DNA coils (dsDNA and ssDNA) the parameters for the statistical-mechanical description of the behavior of these polymers have been determined. The calculations indicate that dsDNA follows mainly the classical Zimm-type kinetics rather than the Rouse one as it was originally proposed. Single-stranded DNA also behaves predominantly as the Zimm polymer. For dsDNA the Kuhn length agrees with the commonly accepted value in the literature while in the case of ssDNA it takes a value much larger than it is usually cited in the literature.

## I. INTRODUCTION

The dynamics of flexible polymers in dilute solution is most often studied within the bead-spring Rouse and Zimm phenomenological models [1-4]. In spite of a great success of these models in the description of the long-time behavior of polymer macromolecules, still a number of problems remains unsolved [5-7]. Moreover, new "puzzles" appear even in the understanding of the polymers that are intensively studied many years [8]. The aim of this paper is to show that some of the problems may be due to an inappropriate use of the Rouse-Zimm (RZ) theory in the interpretation of experimental data. We will consider the recent experiments [8], but believe that our approach could be of more general interest. We come from known facts, which are however often neglected. First, the parameters used in the description of the polymer (modeled by a chain of beads) behavior are phenomenological and separately they have no physical sense; e.g., the bead radius is not a radius of any real object.

Only certain combinations of the parameters have a sense of observable quantities and have to be determined from experiments. Secondly, this determination should be done coming from the joint RZ model, which includes the hydrodynamic interaction between the beads but does not *a priori* neglect the Rouse contribution in the equation of motion for the beads. The joint RZ model thus contains the Rouse ad Zimm models as limiting cases (infinitely small and large draining parameters, respectively). Moreover, in the consideration of the internal modes of polymer relaxation one must take into account that the normal modes have their own draining parameter that depends on the mode number $p = 1, 2, ...$ When $p$ increases, this draining parameter decreases and beginning from some $p$ all the higher internal modes become the Rouse modes even if the motion of the whole polymer is predominantly of Zimm type. And finally, the type of the dynamics depends also on the time. Consider, for example, the mean square displacement (MSD) of a polymer segment. At sufficiently short times its dynamics in a flexible polymer is of Rouse type and with time growing it changes to Zimm type dynamics (the relative contribution of the modes with small draining parameters decreases, or, in other words, the influence of the hydrodynamic interaction increases). This crossover is not possible to describe coming from approximate expressions, according to which the MSD is given by the $t^{1/2}$ and $t^{2/3}$ laws for the "Rouse" and "Zimm" polymers, respectively [9, 10]. We shall demonstrate in this paper that the application of the more fundamental joint Rouse-Zimm model in the interpretation of experiments could lead to essentially different results in comparison with the approach when the continuum of the relaxation modes is assumed, and when the polymer is from the beginning considered as being of Rouse or Zimm type. The recent experimental work [8] allowed us to realize such analysis in a relatively simple way.

The mentioned work [8] together with the experiments [11] represents a significant progress in the use of the fluorescence correlation spectroscopy (FCS) [12] to the study of the dynamics of macromolecules in solution. While in Ref. 11 the general features of the internal dynamics of single randomly labeled DNA in aqueous solution were assessed, the work [8] reports the first FCS measurements of the stochastic motion of individual monomers within isolated polymer coils. Single- and double-stranded DNA (ssDNA and dsDNA) were used as model polymers. A fluorescent label was specifically attached to a single base on the DNA molecules and the motion of the labeled monomer was monitored. One of the advantages of FCS over more classical techniques such as the dynamic light and neutron scattering is in a simpler interpretation of the experiments: the motion of one specifically tagged monomer is, in theory, easier to describe than the complex motions of the whole polymer coil. It is also not necessary to calculate the dynamic structure factor depending by a complicated manner on the scattering wave vector. FCS is thus very suitable for testing the theories of polymer dynamics. In Ref. 8 the time dependence of the MSD of the end monomer has been measured and compared to the Rouse and Zimm models. According to Ref. 8, a rather puzzling agreement with the prediction of the Rouse model for the internal dynamics of polymers, $\langle r^2(t)\rangle \sim \sqrt{t}$, has been observed for dsDNA for a wide range of time scales, while in the case of ssDNA the monomer followed the Zimm-type $\sim t^{2/3}$ kinetics, independent on any polymer parameters. The latter regime corresponds to the common view on the dynamics of flexible polymers. The unexpected observation of the Rouse behavior for dsDNA was explained qualitatively but appears to be in quantitative disagreement with current theories of polymer dynamics.

In the present work the experiments [8] are analyzed in more detail. The analysis is based on the determination of the phenomenological parameters that enter the Rouse-Zimm (RZ) description of the polymer dynamics and should be extracted from experiments. Within the Rouse model for flexible polymers, the hydrodynamic radius of the coil (= $Nb$, where $N$ is the number of beads and $b$ the bead radius) can be chosen as such a parameter. The second unknown quantity in the model is, e.g., the gyration radius $R_G = (Na^2/6)^{1/2}$. Here, $a$ is the mean square distance between the neighboring beads on the chain, assuming that the equilibrium distribution of the beads is Gaussian. The behavior of the pure Zimm polymer is governed by the only phenomenological parameter $R_G$. These parameters determine the relaxation times of the polymer internal normal modes, $\tau_p$, $p = 1, 2,…$, and the diffusion coefficient of the motion of the coil as a whole, $D$. An important quantity is the "draining parameter" $h = 2\sqrt{3N/\pi} b/a$ that indicates whether the hydrodynamic interaction in solution is strong or not, i.e. whether the dynamics is of the Zimm ($h \gg 1$) or Rouse ($h \ll 1$) type. For the internal modes the draining parameter depends on the mode number $p$, $h(p) = \tau_{pR}/\tau_{pZ} = h/\sqrt{p}$, for the diffusion of the polymer as a whole the draining parameter is $D_Z/D_R = 4\sqrt{2}h/3$. Here $\tau_{pi}$ are the relaxation times in the corresponding, Rouse or Zimm, model, and $D_i$ are the diffusion coefficients of the coil [3].

Usually the dynamics of polymers in solution is described by one of these models. However, in general the polymer is not of the pure Zimm or Rouse type and its behavior should necessarily possess more or less expressed features of both the models. As discussed above, it thus seems reasonable to begin the description of the polymer dynamics coming from the more general theory that contains the Rouse and Zimm models as limiting cases [10, 13, 14]. The model parameters should be determined from experiments and only then one can consistently interpret the behavior of the polymer in the studied experimental conditions. We used this approach for the interpretation of the experiments [8]. In particular, fitting the general RZ theory to the experimental data, we have, contrary to the work [8], identified the behavior of dsDNA as being essentially of the Zimm type, with the model parameters very different from those used in the cited work. Also the single-stranded DNA follows mainly the Zimm-type dynamics. We have determined the parameters of this polymer which allows one to calculate two key quantities for the statistical-mechanical description of the universal behavior of polymers in solution: the radius of gyration and the persistence length. The value of the latter parameter has been found much larger than the values usually determined from other experiments.

## II. THE JOINT ROUSE-ZIMM MODEL

For the parameters used in Ref. 8 in the interpretation of the data the draining parameter $h$ does not obey any of the above conditions, i.e. it is neither small nor large. For example, for dsDNA with 23100 base pairs ($N = 68$, $a = 100$ nm, $b = 12.8$ nm) one finds $h \approx 2$ and $D_Z / D_R \approx 4$, so that the observed Rouse dynamics is not expected for flexible polymers. For this reason we have analyzed the experimental data with no preliminary assumption concerning the applicability of a specific dynamics. The used model can be summarized as follows (some remarks on the substantiation of the RZ model are given in Appendix A).

The MSD of a polymer bead was calculated in a number of papers in connection with the evaluation of the dynamic structure factor of the light or neutron scattering from polymers within the bead-spring models [3, 4, 9, 10, 13, 14]. The MSD of the $n$th bead from $N$ beads in the chain has the form

$$\langle r_n^2(t) \rangle = 3\langle [x_n(0) - x_n(t)]^2 \rangle = 6 \sum_{p=0}^{\infty} [\psi_p(0) - \psi_p(t)](4 - 3\delta_{p0}) \cos^2 \frac{\pi n p}{N}, \quad (1)$$

where each of the Cartesian components $x_n$ of the radius vector of the bead is expressed through the normal coordinates $y_p$,

$$x_n(t) = y_0(t) + 2 \sum_{p=1}^{\infty} y_p(t) \cos \frac{\pi n p}{N}, \quad (2)$$

and $\psi_p$ are the correlation functions for the normal modes,

$$\psi_{p \geq 1} = \langle y_p(t) y_p(0) \rangle, \qquad \psi_0(0) - \psi_0(t) = \frac{1}{2} \langle [y_0(t) - y_0(0)]^2 \rangle. \quad (3)$$

It is seen from the inverse Fourier transformation to Eq. (2) that the $y_0$ mode describes the motion of the center of inertia of the coil. Its contribution to the bead MSD is given by $\psi_0(0) - \psi_0(t) = Dt$. The internal modes numbered with $p > 0$ relax exponentially, $\psi_p(t) = \psi_p(0)\exp(-t/\tau_p)$, with $\psi_p(0) = Na^2/(6\pi^2 p^2)$. For the MSD of the end bead, which corresponds to the experiments [8], we thus have from Eq. (1)

$$\langle r^2(t) \rangle = 6Dt + \frac{4Na^2}{\pi^2} \sum_{p=1}^{\infty} \frac{1}{p^2} \left[ 1 - \exp\left( -\frac{t}{\tau_p} \right) \right], \quad (4)$$

where in the RZ theory $D = D_R + D_Z$ is the Kirkwood diffusion coefficient of the polymer chain [3, 7, 14-16]. The relaxation rates [10, 13, 14] can be expressed in the form $1/\tau_p = 1/\tau_{pR} + 1/\tau_{pZ}$. This expression is valid if the mode number $p$ is small compared to the number of beads in the chain [17]. The limiting diffusion coefficients and the relaxation times are given by the known formulas [3]

$$\tau_{pR} = \frac{2N^2 a^2 b \eta}{\pi k_B T p^2}, \qquad \tau_{pZ} = \frac{(N^{1/2} a)^3}{(3\pi p^3)^{1/2}} \frac{\eta}{k_B T}, \tag{5}$$

$$D_Z = \frac{8 k_B T}{3\sqrt{6\pi^3 N} \eta a}, \qquad D_R = \frac{k_B T}{6\pi N b \eta}, \tag{6}$$

where $\eta$ is the viscosity of the solvent. More general calculations that take into account the effects of fluid inertia (the hydrodynamic memory) on the bead motion are given in Ref. 14. Equation (4) represents the long-time limit of the theory [14], $t \gg R^2 \rho/\eta$, where $\rho$ is the solvent density and $R$ the coil radius.

Converting at $t \ll \tau_1$ the sum (4) into the integral, one has in the Rouse limit [9]

$$\langle r^2(t) \rangle \approx 6 D_R t + \frac{a}{\pi} \left( \frac{8 k_B T}{\eta b} t \right)^{1/2}, \tag{7}$$

and for the Zimm case (correcting the numerical factor of the result [10] for the internal modes)

$$\langle r^2(t) \rangle \approx 6 D_Z t + \frac{4}{\pi^2} (3\pi)^{1/3} \Gamma\left(\frac{1}{3}\right) \left( \frac{k_B T}{\eta} t \right)^{2/3} \approx 6 D_Z t + 2.3 \left( \frac{k_B T}{\eta} t \right)^{2/3}. \tag{8}$$

In Ref. 8, Eqs. (7) and (8) were used (without the diffusion terms and with the factor 2 instead of 2.3) in the description of the experiments. In the case of dsDNA it was concluded that for a wide range of time scales (from 0.02 to 10 ms) the behavior of the polymer is surprisingly of the Rouse type, and for longer times it follows universal Zimm kinetics. It should be however noted that Eqs. (7) and (8) are only crude approximations to Eq. (4) with incorrect time dependence for both long and short times. In particular, within the interval 0.02 – 10 ms Eq. (7) cannot be used at all: taking the polymer parameters from Ref. 8, the calculations show that the sum in Eq. (4) constitutes from about 8 to 80% of its approximation given by the second term in Eq. (7). In Ref. 8, the "Rouse" behavior was qualitatively explained estimating how the internal modes with $p > h^2$ contribute to the observed MSD. Let us consider this explanation in more detail. For $p > h^2$ the draining parameter becomes small and the dynamics of such modes should be of Rouse type. In the experiment [8] the Rouse dynamics can be expected for the modes with $p > 4$ when $h(p) < 1$. We shall call these modes the Rouse modes even if the condition $h(p) \ll 1$ is not well satisfied (e.g. for $p = 5$ we have $h(5) \approx 0.9$). The estimation of the contribution of such modes to the MSD shows that the first four modes contribute almost 87 per cent to the total amplitude $\langle r^2(\infty) \rangle$ determined by the internal modes.

Only about 13 % can be assigned to the "Rouse" modes. The time-dependent contribution of the Rouse modes relative to the total MSD due to the internal modes, $F_R(t)$, is shown on Fig. 1. The calculations were done according to the formulas

$$F_R(t) = [f_\infty(t) - f_{h^2}(t)] / f_\infty(t), \qquad f_{h^2}(t) = \sum_{p=1}^{h^2} p^{-2}[1 - \exp(-t/\tau_p)] \qquad (9)$$

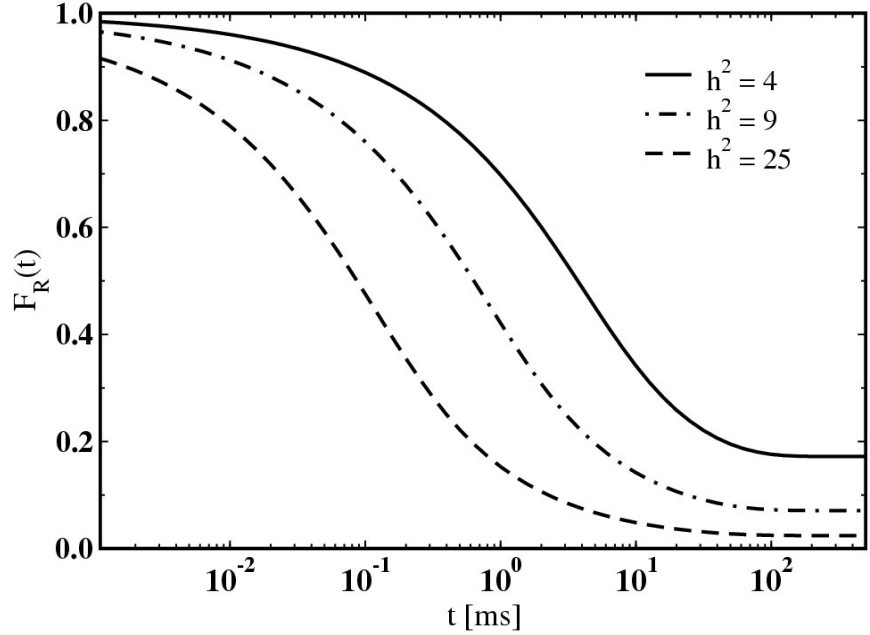

**FIG. 1.** Time dependence of the contribution of the internal "Rouse" modes with $p > h^2$ ($h$ is the draining parameter for the lowest Rouse mode) relative to the total MSD due to the internal motion, calculated from Eq. (9).

The upper line corresponds to the modes with $h(p) < 1$ ($p > 4$), the lowest line is for $h(p) < 0.4$ ($p > 25$). In the time region 0.02 to 10 ms (where according to Ref. 8 the behavior of the polymer is of Rouse type) the contribution of the Rouse modes changes from approximately 95% to 34% (at $t = 1$ ms being 70 %) of the total internal part of the MSD, and at 0.2 s the Rouse contribution falls to 17 %. Qualitatively it agrees with the explanation of the observation in Ref. 8 but quantitative differences from the pure Rouse or Zimm behavior are evident. Moreover, estimations show that also the diffusion contribution plays a significant role. For 23100 bp (base pairs) DNA, the temperature $T = 293$ K, the solvent viscosity $\eta = 1$ mPa s, and $Na^2 = 0.68$ $\mu m^2$ [8], the Zimm diffusion coefficient is $\approx 0.96$ $\mu m^2$/s. Consequently, in the time scales from 10 to 200 ms (where according to Ref. 8 the Zimm behavior is observed), the diffusion term $6D_Z t$ changes from 0.06 do 1.2 $\mu m^2$, while the contribution from the internal modes produces according to Eq. (4) from 0.27 to 2 $\mu m^2$. That is, the diffusion term contributes from 20 to almost 60 per cent and cannot be neglected. For shorter DNAs the importance of this term becomes even more significant.

It is thus seen from Fig. 1 and the above consideration that the question about which type of dynamics is dominant for a given set of the polymer parameters can be solved only comparing the MSD for the pure Rouse and Zimm models with the MSD as it follows from the joint RZ model with the same parameters. Such a comparison is shown on Fig. 2. The parameters for numerical calculations using Eqs. (4 – 6) are the same as those used in Ref. 8 for the 23100 bp DNA. It is seen that none of the models corresponds to the experimental data in a notable time scale. When the diffusion term is omitted as in Ref. 8, the correspondence with the Zimm model (using Eq. (4)) is not observed in the region from 10 to 200 ms where according to Ref. 8 the experiment agrees with the approximate result (8) without the diffusion term. An agreement with the Zimm model is seen only in a narrow region around 10 ms. At "short" times the Rouse model is close to the experiment only at the times of tens ms. It can be concluded that for the parameters used in Ref. 8 the correspondence between the data and the theory is unsatisfactory for all the three models. Moreover, neither the Rouse nor the Zimm model corresponds to the more general RZ model. This indicates that the correct polymer parameters should be different.

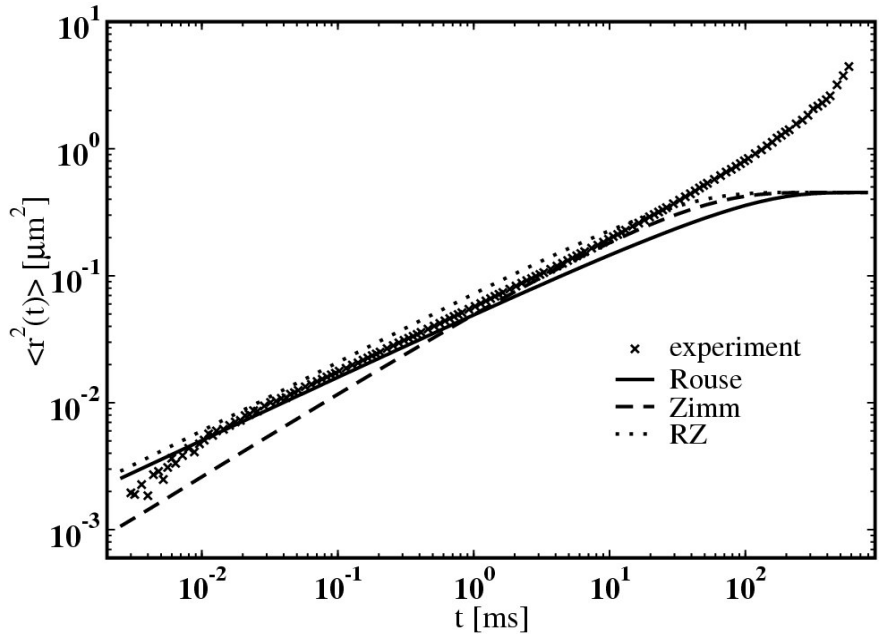

**FIG. 2.** MSD of the end bead from Eq. (4) (the diffusion term is omitted). The experimental data [8] on dsDNA in aqueous solution ($T$ = 293 K and $\eta$ = 1 mPa s) are compared to the Rouse, Zimm, and joint RZ models with the parameters $N$ = 68, $a$ = 100 nm, and $b$ = 12.8 nm [8].

### III. RESULTS AND DISCUSSION

The polymer parameters have been obtained by fitting the RZ model (4) to the experimental data on ds and ssDNA of various lengths [8]. In accordance with the above discussion, the diffusion term is kept in the consideration. The minimization of the root mean squares between the theoretical and experimental curves has been performed by simulating annealing technique [18]. Below are examples of such calculations.

*Double-stranded DNA*

Here we give the results for the longest studied dsDNA containing 23100 base pairs. Although dsDNA is known to be a semi-flexible polymer, the statistics and dynamics of polymer conformation at the length scales larger than the Kuhn length $l$ should be close to those of the flexible coils. As in Ref. 8, we thus look at the dynamics of dsDNA using the theory for flexible polymers. The optimization to the data was done with the natural conditions $Na < L$ and $b < a/2$. The contour length of the polymer $L$ can be calculated using the known distance between the base pairs along the chain (0.34 nm). Distributing the beads in such a way that the distance between the neighboring beads is 100 nm as in Ref. 8, we obtain $N = 78$. This number was fixed in the calculations. Leaving the parameters $a$ and $b$ free (except the above limitations), we optimized the RZ model to the experimental data. The result is shown on Fig. 3. From the optimization the parameters $a = 99.1$ nm, $b = 49.5$ nm, and $h \approx 8.6$ have been obtained that indicates the approximate applicability of the Zimm model.

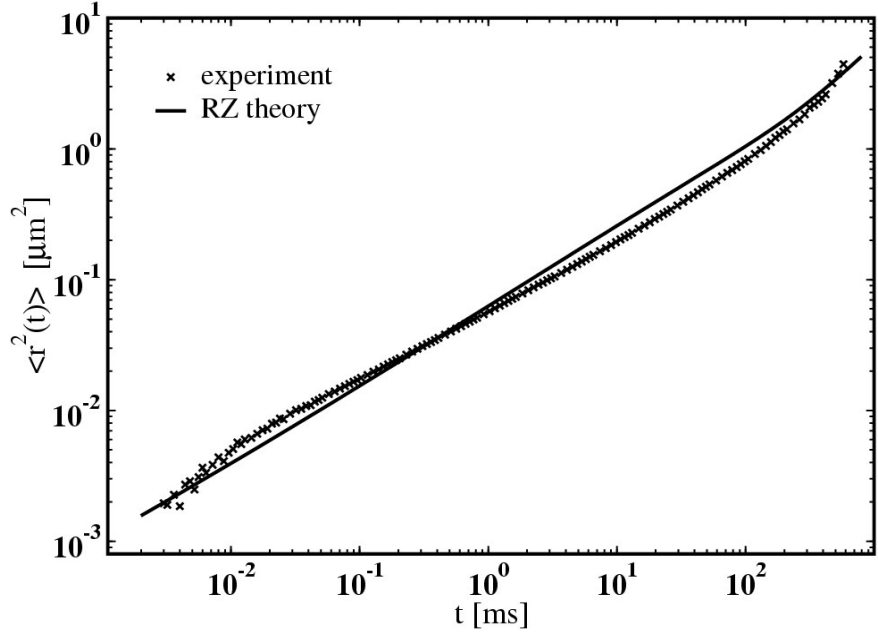

**FIG. 3.** MSD from Eq. (4) for the RZ model optimized to the experimental data [8] for dsDNA (23100 bp) in aqueous solution at $T = 293$ K and $\eta = 1$ mPa s. The optimization yielded the polymer parameters $N = 78$, $a = 99.1$ nm, and $b = 49.5$ nm with the draining parameter $h \approx 8.6$ showing the predominant Zimm-type dynamics.

This was confirmed by the fits of the pure Zimm model that yielded close parameters. When we optimized the pure Rouse model to the data, the obtained parameters led to the Kuhn length (found from its definition $Na^2 = Ll$) which was too small – only about 30 nm while the commonly accepted value in the literature is about 100 nm, in accordance with the solution for the RZ case ($l \approx 98$ nm). This is an additional argument against the validity of the pure Rouse model for the description of the dsDNA dynamics. In summary, as distinct from the conclusion in Ref. 8, the RZ (being predominantly Zimm) model should be preferred. The discrepancies with the experiment seen on Fig. 3 could be possibly ascribed to a semi-flexible nature of dsDNA.

*Single-stranded DNA*

As distinct from dsDNA for which the Kuhn length can be considered as a relatively known quantity not very different from 100 nm, the Kuhn length of ssDNA is sequence and solvent dependent and varies in a wide range, to our knowledge usually from about 1 to 4 nm (e.g. 0.8 nm [19, 20], 1.4 nm [21, 22], 4 nm [4, 23]), in Ref. 24 the value 10 nm is given. For a discussion see Refs. 23, 24. Since $l$ from the experiments [8] is unknown, we considered it as a free parameter to be determined in the optimization procedure. It was also required that $a$ is smaller than the distance between the beads along the chain (this means that $a > (1 - 1/N)l$) and larger than the bead diameter.

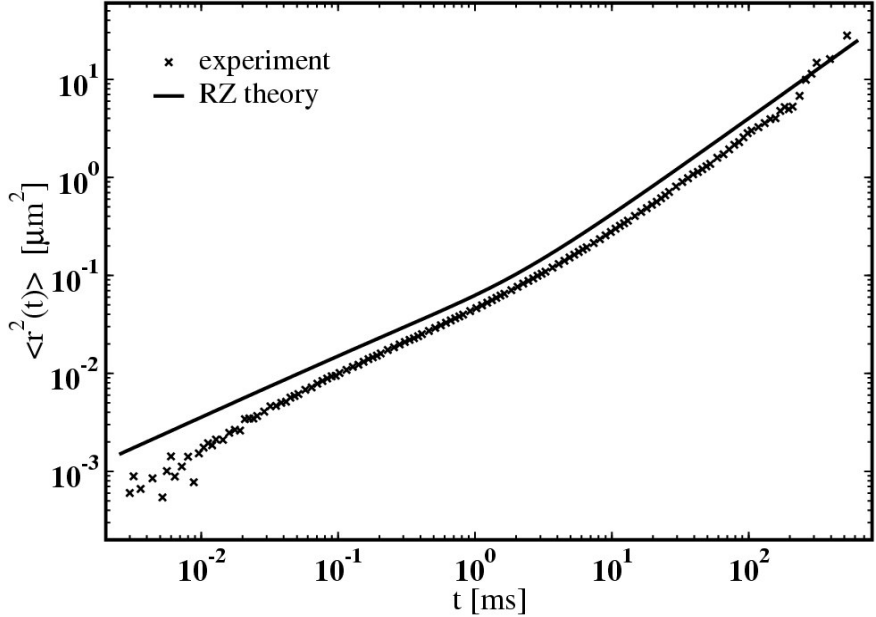

**FIG. 4.** MSD from Eq. (4) for the RZ model optimized to the experimental data [8] for ssDNA (6700 bases) at $T = 310$ K and $\eta = 0.69$ mPa s. The polymer parameters are $a = 9.15$ nm, $b = 4.56$ nm, and $N = 422$. The draining parameter $h \approx 20$ indicates the Zimm-type dynamics.

An example of the optimization of the MSD to the ssDNA data is shown on Fig. 4 (6700 bases, $T = 310$ K, the solvent viscosity $\eta = 0.69$ mPa s, the distance between the bases 0.58 nm [25]. The length of such DNA is $L = 3886$ nm. The optimization then yielded the following parameters: $a = 9.15$ nm, $b = 4.56$ nm and $N = 422$. This corresponds to $h \approx 20$ so that the polymer follows essentially the Zimm dynamics (if the pure Zimm regime is assumed from the beginning, the optimization gives slightly different values $N = 414$ and $a = 9$ nm). Note that for any set of the model parameters the experimental MSD values are lower than the theoretical ones; this resembles the well-known long-lasting "puzzle" between the Zimm theory and dynamic scattering experiments [5]. The discrepancy with the data at the shortest time scales (several $\mu s$) can be partially assigned to processes related to the internal dynamics of the fluorophore only [8].

## IV. CONCLUSION

Within the bead-spring models, the experiments (such as the dynamic light and neutron scattering) on polymers are usually described coming from the Rouse or Zimm theory of polymer dynamics. We have demonstrated in this paper that such approach can lead to an incorrect determination of the polymer phenomenological parameters and to a misleading interpretation of its behavior. In a more correct approach the joint theory should be used in which the Rouse and Zimm models are just limiting cases of infinitely small or large draining parameter. If this parameter is finite, one must be careful with the identification of the polymer behavior as being of the Rouse or Zimm type. First of all this concerns the relaxation of the internal normal modes of the polymer since their draining parameter depends on the mode number. The second problem considered in this paper concerns the often used assumption of the continuous distribution of the internal modes. This approximation is valid only in a restricted time interval, in real situations often fails, and its use can give an improper identification of the polymer dynamics. Moreover, within the joint RZ model the type of the dynamics depends on the time: every flexible polymer changes its behavior from the Rouse dynamics at short times to the Zimm one at longer times. This interesting feature is lost when a less general theory is used or if the internal modes are assumed to be continuously distributed.

We have used the joint RZ theory with the discretely distributed normal modes to analyze the recent experiments in which the kinetics of individual monomers within the polymer coil was observed using the FCS technique [8]. Such analysis is of self-dependent interest also because an unexpected behavior of double-stranded DNA has been reported in these experiments. We have demonstrated that the parameters used in Ref. 8 do not correspond to the experimental data. Optimizing the theory to the data on double- and single-stranded DNA, new sets of their parameters have been obtained. The calculations indicate that dsDNA kinetics is mainly of the Zimm type rather than the Rouse one [8], and the values of its parameters essentially differ from those in the original work. The ssDNA also behaves predominantly as the Zimm polymer. From the determined parameters two key parameters for the statistical-mechanical description of the universal behavior of flexible polymers in solution can be obtained - the gyration radius and the persistence length. For the presented results on dsDNA the Kuhn length agrees with the commonly accepted value. For ssDNA the found Kuhn length is about 9.1 nm, a value larger than the values usually cited in the literature. It should be noted that the proposed description of the experiments [8] cannot be considered as a confirmation of the RZ theory for the studied polymers. There are discrepancies between the theory and experiment that cannot be explained within the used model. In the case of dsDNA the differences could be probably partially assigned to the semi-flexible nature of the polymer. Single-stranded DNA is however considered a flexible polymer so that the RZ theory should be applicable, but the experimental data for the MSD always lie below the theoretical calculations. We tried to explain this problem taking into account the finite resolution time in the FCS experiments (about 1 $\mu$s) and a possible influence of the inertial effects during the bead motion in the solvent at short times [14, 26], but these effects do not notably improve the consideration. In spite of the long-standing investigations of the dynamics of polymers in solution, even the simplest case of the dynamics of individual long flexible polymers still represents a challenge in the theory of polymer dynamics.

## ACKNOWLEDGMENTS

We are greatly indebted to O. Krichevsky for providing us with the experimental data [8]. This work was supported by the Marie Curie European Reintegration Grant MERG-CT-2004-506291 within the EC Sixth Framework Program, and by the grant VEGA 1/0429/03 from the Scientific Grant Agency of the Slovak Republic.

## APPENDIX: REMARKS ON THE JOINT ROUSE-ZIMM MODEL

The basic equation of the Zimm model of polymer dynamics is the equation of motion for the position vector of the $n$th bead [2 - 4],

$$\frac{d\vec{x}_n}{dt} = \frac{1}{\xi}\left(\vec{f}_n^{ch} + \vec{f}_n\right) + \vec{v}(\vec{x}_n). \qquad (A.1)$$

(The inertial term is omitted.) Here, $\vec{f}_n^{ch}$ is the force with which the neighboring beads act on the $n$th bead, $\vec{f}_n$ is the random force due to the motion of the molecules of solvent, $\vec{v}(\vec{x}_n)$ is the velocity of the solvent in the place of the $n$th bead due to the motion of other beads, and $\xi$ is the friction coefficient (for a spherical particle $\xi = 6\pi\eta b$, where $\eta$ is the solvent viscosity and $b$ is the bead radius. As distinct from the theory by Rouse [1, 4], where the solvent is nonmoving ($\vec{v} = 0$), Eq. (A. 1) takes into account the hydrodynamic interaction [3, 4]. Within the Zimm theory the velocity field $\vec{v}(\vec{x}_n)$ is expressed through the Oseen tensor $\hat{H}$,

$$\frac{d\vec{x}_n}{dt} = \frac{1}{\xi}\left(\vec{f}_n^{ch} + \vec{f}_n\right) + \sum_{m \neq n} \hat{H}(\vec{x}_n - \vec{x}_m)\left(\vec{f}_m^{ch} + \vec{f}_m\right). \qquad (A.2)$$

In the sum $m \neq n$ since the velocity field in the point $n$ is created by all other beads except the $n$th one. If one, following [4], formally defines the Oseen tensor for $n = m$ as $\hat{H}_{nn} = \delta_{nm}/\xi$, the summation in Eq. (A. 2) can be extended to all numbers $m$ from 0 to $N$ (the number of beads in the chain). Usually, the continuum approximation with respect to the variable $n$ is used: $\sum_m (...) \to \int_0^N (...)dn$. After this step, however, the first term on the right in Eq. (A. 1) completely disappears. It does not matter what was the constant $\xi$; it does not influence the solution at all. One could expect that the Zimm model should generalize the earlier Rouse model. If we act as described above, this is not the case: the two models are independent. Moreover, imagine that in the Rouse model (with $\vec{v} = 0$), that is, with the basic equation

$$\frac{d\vec{x}_n}{dt} = \frac{1}{\xi}\left(\vec{f}_n^{ch} + \vec{f}_n\right), \qquad (A.3)$$

we lose in the continuum approximation the term on the right; the model would become meaningless.

If we want to obtain a generalization of the Rouse model, we have in the continuum approximation to keep the first term on the right in Eq. (A. 2). The $m = n$ term in the sum can be defined arbitrarily: it will not influence the result of integration. Thus, the Rouse-Zimm equation in the continuum limit ($\vec{x}_n(t) \to \vec{x}(t,n)$ and so on) should be

$$\frac{\partial \vec{x}(t,n)}{\partial t} = \frac{1}{\xi}\left[\vec{f}^{ch}(t,n) + \vec{f}(t,n)\right] + \int_0^N dm \hat{H}(n,m)\left[\vec{f}^{ch}(t,m) + \vec{f}(t,m)\right]. \qquad (A.4)$$

As discussed in the main text, depending on the draining parameter the Rouse-Zimm model gives the description of the polymer behavior more close to the Rouse or Zimm dynamics. In general, however, both terms on the right side of Eq. (A.4) should be kept in consideration. Then one can follow the standard way described in detail e.g. in the monograph [4]. This will lead, in particular, to Eqs. (1) - (6). The model used in this paper is valid for steady motion of the beads and solvent. A more general Rouse-Zimm theory that takes into account inertial effects in the polymer motion is described in Ref. 14.

___________________________________________________________